\documentclass[12pt]{article}
\usepackage[english]{babel}
\usepackage{hyperref,hypcap,ae,tikz}
\usepackage{graphicx}
\usepackage{bm}
\usepackage{amsmath}
\usepackage{amsfonts}
\usepackage[T1]{fontenc}
\usepackage{textcomp}
\usepackage{xcolor}
\usepackage{lmodern}
\usepackage[caption=false]{subfig}

\hypersetup{
	colorlinks=true,
	citecolor=blue,
    linkcolor=magenta,
	}
\begin{document}

\title{{\bf Arrow of time and gravitational entropy in collapse }}

\author{\bf{Samarjit Chakraborty $^1$, Sunil D. Maharaj $^2$,}
\\{\bf Sarbari Guha$^1$\footnote{srayguha@yahoo.com} and Rituparno Goswami $^2$} \\ $^1$ Department of Physics \\ St.Xavier's College (Autonomous), Kolkata 700016, India \\
 $^2$ Astrophysics and Cosmology Research Unit, \\ School of Mathematics, Statistics and Computer Science, \\ University of KwaZulu-Natal, \\ Private Bag X54001, Durban 4000, South Africa}

\maketitle
\section*{Abstract}
We investigate the status of the gravitational arrow of time in the case of a spherical collapse of a fluid that conducts heat and radiates energy. In particular, we examine the results obtained by W. B. Bonnor in his 1985 paper where he found that the gravitational arrow of time was opposite to the thermodynamic arrow of time. The measure of gravitational epoch function $P$ used by Bonnor was given by the ratio of the Weyl square to the Ricci square. In this paper, we have assumed the measure of gravitational entropy $P_{1}$ to be given by the ratio of the Weyl scalar to the Kretschmann scalar. Our analysis indicates that Bonnor's result seems to be validated, i.e., the gravitational arrow and the thermodynamic arrow of time point in opposite directions. This strengthens the opinion that the Weyl proposal of gravitational entropy applies only to the universe as a whole (provided that we exclude the white holes).

\bigskip

KEYWORDS: Gravitational entropy, Gravitational collapse, Arrow of time.

\section{Introduction}
Gravitational entropy (GE) is the entropy associated with the degrees of freedom of the free gravitational field. It was first proposed by Roger Penrose to solve the entropy problem of the initial universe. If the universe is to satisfy the second law of thermodynamics, then the net entropy must increase monotonically from its very inception. However, we know that the cosmic microwave radiation (CMBR) is a near perfect black body radiation indicating that the universe was in thermal equilibrium at the very beginning. This implies that the entropy of the present universe was a maximum at its birth. This contradicts the basic content of the second law of thermodynamics. If the entropy had a maximum value at the beginning, then the entropy of the universe cannot increase any further, thereby leading to a paradox. Another purpose of the proposal of GE was to put the black hole (BH) entropy in a proper context, as it is the only entropy of a system which depends on its area unlike other thermodynamic entropies which are dependent on the volume of the system. Therefore Penrose \cite{Penrose1,Penrose2} proposed the idea of GE, effectively introducing a new entropy associated with the free gravitational field so that the combination of the initial entropy corresponding to the matter content of the universe plus the GE becomes minimum, although the matter part was in thermal equilibrium in the initial universe. As time progresses the total entropy of the universe increases monotonically \cite{Bolejko}. The whole idea was to solve the entropy problem in the universe as we have explained above. This approach also implies the existence of a gravitational arrow of time. The gravitational entropy indicates that in the course of time, the gravitational tidal forces condense matter in the universe, leading to an increasing inhomogeneity, thus giving rise to structure formation. Consequently, it ensures that this GE, measured by some function of the Weyl curvature scalar, reflects the degree of clumping of matter and hence the increase of inhomogeneity in the universe, i.e., the entropy of the universe is mostly carried by the free gravitational field.

The study of gravitational entropy can be performed locally even on astrophysical objects like black holes, wormholes and stars, although the initial motivation was to study entropy growth of the universe globally. As mentioned earlier, the concentration of matter in pockets directly solves the issue of BH entropy, and as we know that most of the entropy in the universe are due to black holes, the BH entropy leads to extreme special cases for gravitational entropy on a local scale. The GE plays a very important role even in the thermodynamics of spacetimes. It gives an overall energetic picture of a region of spacetime along with the gravitational effect of specific geometries. Thus, GE captures the essence of gravitational theories in the form of a thermodynamic tool relating geometry to energy and vice versa.

The cosmic evolution not only gives a sense of time from the gravitational perspective, but also provides an arrow originating at the inception of the universe. In order to determine the gravitational entropy one needs to define the gravitational epoch function, and there are many proposals for this function. In this paper, we undertake this study from the perspective of  Bonnor \cite{Bonnor1,Bonnor2,Bonnor3}, where he compared the gravitational arrow of time with the thermodynamic arrow of time. Bonnor used the gravitational epoch function $ P $, defined as
 \begin{equation}\label{P}
P={(C_{abcd}C^{abcd})}{(R_{ab}R^{ab})}^{-1},
\end{equation}
where $C_{abcd}  $ is the Weyl tensor, and $R_{ab}$ is the Ricci tensor.
Here we will extend his work by using the gravitational
entropy function, proposed by Rudjord et al \cite{entropy1,entropy2}. For our case the gravitational epoch function $P_{1}$ is given by
\begin{equation} \label{P1}
P_{1}={(C_{abcd}C^{abcd})}{(R_{abcd}R^{abcd})}^{-1},
\end{equation}
where $R_{abcd} $ represents the Riemann tensor, which contains both the trace part (Ricci tensor) and the traceless part (Weyl tensor). According to the Weyl curvature hypothesis, the Ricci part dominated over the Weyl part during the initial phase of the universe.

It should be noted that the function $P$ is well behaved in non-vacuum models but fails in empty space. This is due to the fact that for vacuum solutions, the Ricci tensor vanishes, making the function $ P $ unusable. As an example, for external Schwarzschild spacetime, the Ricci tensor vanishes, and we cannot use $ P $. This is why a different proposal $ P_{1} $ was considered by Rudjord et al. in \cite{entropy1}. Not only that $P_{1}$ doesn't leave behind any information which $ P $ can capture, but in addition, also contains the contribution from the Ricci scalar, i.e., how the spacetime differs from Euclidean space. This is important for non-vacuum spacetimes, and the two proposals are different because $ P_{1}^{-1}-2(P)^{-1}=1-1/3(R^2/W) $, where $ W= C_{abcd}C^{abcd}$. Further, it is better to have the Kretschmann Scalar in the denominator, as it not only captures the entire dynamics of the spacetime but it is also singularity sensitive, as it ignores the coordinate singularity in the Schwarzschild spacetime, only diverging at the true central singularity.
As the gravitational epoch function determines the magnitude of the gravitational entropy, the gravitational epoch function is expected to increase monotonically. The entire proposal of gravitational entropy mainly concerns itself with the issue of cosmological entropy, and it requires the absence of any white whole in the beginning of the universe \cite{Bonnor1,Bonnor2,Bonnor3}. \\
Both $P$ and $P_{1}$ are identically zero in the Friedmann-Robertson-Walker models (for all conformally flat spacetimes) of standard cosmology, as the Weyl tensor vanishes in these models. However, $P_1$ is a phenomenological proposal derived by matching the description of the entropy of a black hole \cite{entropy1} with the Hawking-Bekenstein entropy \cite{Bekenstein,SWH1}, that helped Rudjord et al \cite{entropy1,entropy2} to calculate the entropy of Schwarzschild black holes and the Schwarzschild-de-Sitter spacetime. There have been many studies where the gravitational entropy is determined for local astrophysical objects and are shown to be valid, e.g., the Hawking-Bekenstein entropy of a black hole \cite{Chandra}, can be obtained from the gravitational entropy. Such studies have also been conducted by the authors in \cite{gc,cgg,cgg1} using the Weyl proposal due to Rudjord et al \cite{entropy1,entropy2} and the CET proposal \cite{CET}.

\bigskip

In what follows, we test the Rudjord proposal of gravitational entropy from the perspective of the gravitational arrow of time. We have already tested this proposal for different kinds of accelerating black holes and wormholes.
Extensive gravitational entropy studies on LTB dust collapse and cosmic voids were completed by Sussman et al. \cite{suss1,suss2,suss3}. Sussman \cite{suss1} introduced a weighed scalar average formalism (the “q-average” formalism) in the study of spherically symmetric LTB dust models and applied this formalism to the definition of the GE functional proposed by Hosoya et al.\cite{HB}. Subsequently, Sussman and Larena \cite{suss2} analyzed the generic LTB dust models using the CET proposal \cite{CET} and the HB proposal \cite{HB}, and established that the notion of gravitational entropy is a theoretically robust concept which can also be applied to generic spacetimes. They also studied the evolution of the CET gravitational entropy for local expanding CDM cosmic voids employing a nonperturbative approach \cite{suss3}. Another important work was carried out on the analysis of gravitational entropies associated with exotic wormholes and galactic halos  by Lima et al. \cite{LNP}. Recently a detailed study have been conducted on structure formation using the CET proposal in Szekeres class I models by Pizaña et al. \cite{Piza}.

In this study we consider the collapse of a radiating star. The idea is to examine whether in this process the gravitational arrow of time coincides with the well established radiation (thermodynamic) arrow, as the radiation emanating from an object always moves outwards, giving us a clear sense of the arrow of time. The study of such a system is important to check whether the gravitational arrow of time makes sense locally in such cases.

\section{The model}
For a collapsing radiating star we can describe the system in two parts \cite{santos,olivera1,olivera2}, the interior metric representing a collapsing sphere of fluid with radial heat flow and the exterior metric given by the Vaidya metric, which gives us the exterior of the star with outward radiation. This provides us a sense of the thermodynamic arrow of time. We neglect the inward Vaidya solution as it is not applicable here.
The interior of this system can be described by the Einstein field equations with heat flow
\begin{equation}\label{ein}
R_{ab}-\frac{1}{2}g_{ab}R=G_{ab}= T_{ab};\;\;\;\; T_{ab}=8\pi[(\mu + p)v_{a}v_{b}+p g_{ab}+q_{a}v_{b}+q_{b}v_{a}].
\end{equation}
This represents a fluid with energy density $ \mu $, pressure $ p $, heat flow $ q^{a}=(0,q,0,0) $ and four velocity $ v_{a} $. The interior of the star is represented by a spatially isotropic spherically symmetric metric with shear-free motion of the fluid. The spacetime is given by
\begin{equation}\label{int}
ds_{-}^2=-A(r,t)^2 dt^2+B(r,t)^2[dr^2+r^2(d\theta^2+\sin^2\theta d\phi^2)].
\end{equation}
Therefore, we can take the four velocity of the fluid in comoving coordinates as $v^{\alpha}=1/A \delta_{0}^{\alpha}$.
We want to mention that we have retained Bonnor's notation \cite{Bonnor3} in most of the places, as we are examining his work from the perspective of the Rudjord proposal of gravitational entropy.
The exterior metric represents an outgoing radiation field around a spherically symmetric star which can be assumed to be the Vaidya metric as in
\begin{equation}\label{ext}
ds_{+}^2=-\left(1-\frac{2m(u)}{R}\right)du^2-2dudR+ R^2(d\theta^2 +\sin^2\theta d\phi^2),
\end{equation}
where $m$ is the gravitational mass which is a function of the retarded time $u$, and $R$ is the radial coordinate. For the outgoing radiation fluid, the Einstein tensor is given by
\begin{equation}
G_{ab}=R_{ab}=8\pi T_{ab}=-\frac{2}{R^2}\delta_{a}^{0}\delta_{b}^{0}(dm/du).
\end{equation}
Here the Ricci tensor has only one non-zero component $ R_{uu}=-2m(u)_{,u}/R^2 $ and the Ricci scalar vanishes.
For the energy density of the radiation to be positive we must have $ dm/du<0 $, i.e., the central object is losing mass because of the outgoing radiation.
\\
The exact solution of the interior metric with the field equation can be taken as
\begin{equation}\label{int_soln}
A=A_{0}(r),\;\; B=B_{0}(r)f(t);\;\; f(t)>0,\;\; A_{0}>0,\;\; B_{0}>0 ,
\end{equation}
where $ A_{0}, \, B_{0} $ are solutions of the interior metric for the static perfect fluid without heat flow, having $ \mu_{0} $ as the energy density and $ p_{0} $ as the isotropic pressure. Further, $ f(t) $ is a positive function to be determined \cite{Bonnor3}. A detailed study on heat conducting shear-free fluid solutions was done by Sussman in \cite{suss0}.
For recent treatments of the geometrical properties of the metric \eqref{int_soln} in the context of the radiating stars see Paliathanasis et al \cite{X1} and Ivanov \cite{X2}. \\
The non-zero components of the field equation are given by the following set:
\begin{eqnarray}\label{einc}
G_{00}^{-}=8\pi \mu A^2, G_{11}^{-}=8\pi p B^2, G_{22}^{-}=\frac{1}{\sin^2\theta}G_{33}^{-}=8\pi p B^2 r^2, G_{01}^{-}=-8\pi q B^2 A.
\end{eqnarray}
Here the components of the Einstein tensor are functions of $A$ and $B$. From \eqref{einc}, it follows that the isotropic pressure $ p $, energy density $ \mu $ and heat flow $ q_{a} $ for the interior spacetime \eqref{int} with the conditions \eqref{int_soln} are given by the equations
\begin{equation}\label{pr}
p=\frac{p_{0}}{f^2}-\frac{1}{8\pi A_{0}^2}\bigg\lbrace2\frac{\ddot{f}}{f}+\left(\frac{\dot{f}}{f}\right)^2\bigg\rbrace ,
\end{equation}
\begin{equation}\label{mu}
\mu=\frac{\mu_{0}}{f^2}+\frac{3}{8\pi A_{0}^2}\left(\frac{\dot{f}}{f}\right)^2 ,
\end{equation}
\begin{equation}\label{qa}
q^{\alpha}=q\delta_{1}^{\alpha}=-(4\pi A_{0}^2 B_{0}^2 f^3 )^{-1}A_{0}^{'}\dot{f}\delta_{1}^{\alpha} ,
\end{equation}
where the energy density and isotropic pressure for the static solution (without heat flow) are given by $ \mu_{0} $ and $ p_{0} $ respectively. There is no heat flow for the static solution.

\section{Matching conditions and consequences}
Let us consider a spherical hypersurface $\Sigma $, which is a timelike three space dividing the entire spacetime into two distinct four dimensional manifolds denoted by $\mathcal{S}^{-}$ and $ \mathcal{S}^{+} $, denoting the interior and exterior region respectively. The intrinsic metric of $\Sigma$ in comoving coordinates is
\begin{equation}
ds_{\Sigma}^2=-d\tau^2+\mathcal{R}^2(\tau)(d\theta^2 +\sin^2\theta d\phi^2).
\end{equation}
Taking the coordinates of the spacetime $ \mathcal{S}^{-} $ we can get the equation of the boundary surface in the form $ f(r,t)=r-r_{\Sigma}=0 $, where $ r_{\Sigma} $ is a constant. In this coordinate the unit normal vector to $\Sigma$ is
\begin{equation}
n_{\alpha}^{-}=\lbrace0,B(r_{\Sigma},t),0,0\rbrace .
\end{equation}
We now match the $ds_{\Sigma}^2=(ds_{-}^2)_{\Sigma}$ at $ r=r_{\Sigma} $ which gives us $ A(r_{\Sigma},t)dt/d\tau=1 $ and $r_{\Sigma}B(r_{\Sigma},t)= \mathcal{R}(\tau) $. Moreover we can calculate the extrinsic curvature $(K_{ij}^{-})$ interior to the hypersurface $ \Sigma $. The non-zero components are given by the following:
\begin{eqnarray}
K_{\tau\tau}^{-}=\left(-\frac{1}{AB} \frac{\partial A}{\partial r}\right)_{\Sigma}, \quad K_{\theta\theta}^{-}=\frac{1}{\sin^2\theta}, \quad K_{\phi\phi}^{-}=\left(r\frac{\partial(Br)}{r}\right)_{\Sigma}.
\end{eqnarray}
Next we consider the exterior spacetime $ \mathcal{S}^{+} $. The equation for the surface $\Sigma$ becomes $ f(R,u)=R-R_{\Sigma}(u)=0 $. The normal unit vector to the boundary surface is then given by
\begin{equation}
n_{\alpha}^{+}=(2dR_{\Sigma}/du+1-2m/R_{\Sigma})^{-1/2}(-dR_{\Sigma}/du,1,0,0).
\end{equation}
From the matching of the two metrics we get $ R_{\Sigma}(u)=\mathcal{R}(\tau) $, and $$(2dR/du+1-2m/R)_{\Sigma}=(1/(du/d\tau)^2)_{\Sigma}  .$$
Therefore we can rewrite the normal unit vector as $ n_{\alpha}^{+}=(-\tilde{R},\tilde{u},0,0) $, where $ \tilde{X} $ represents $ dX/d\tau $. Finally we can write the non-zero components of the extrinsic curvature $ K_{ij}^{+} $ for $ \Sigma $ as
\begin{eqnarray}
K_{\tau\tau}^{+}=\left(\frac{\tilde{\tilde{u}}}{\tilde{u}}-\tilde{u}m/R^2\right)_{\Sigma}, \, K_{\theta\theta}^{+}=\frac{1}{\sin^2\theta}, \, K_{\phi\phi}^{+}=R(\tilde{u}(1-2m/R)+\tilde{R})_{\Sigma} .
\end{eqnarray}
Now using the second junction condition $ [K_{ij}]=K_{ij}^{+}-K_{ij}^{-}=0 $, and the angular components of the extrinsic curvatures at $ \Sigma $, we get
\begin{equation}\label{anK}
\bigg\lbrace r \dfrac{\partial(Br)}{\partial r}\bigg\rbrace_{\Sigma}=\lbrace R(\tilde{u}(1-2m/R)+\tilde{R})\rbrace_{\Sigma} .
\end{equation}
From the first junction condition for the matching of the two metrics together with \eqref{anK}, we get the expression for the mass function $m(u)$ as
\begin{equation}\label{sigmass}
m(u)=\bigg\lbrace \frac{r^3 B}{2A^2}(\partial B/\partial t)^2-r^2 (\partial B/\partial r) - \frac{r^3}{2B}(\partial B/\partial r)^2 \bigg\rbrace_{\Sigma} .
\end{equation}
This is the total energy trapped inside the hypersurface $ \Sigma $. Now we use the second junction condition but this time with the $ \tau $ components, to arrive at the following relation
\begin{equation}\label{junt}
\left(-\frac{1}{AB} \frac{\partial A}{\partial r}\right)_{\Sigma}=\left(\frac{\tilde{\tilde{u}}}{\tilde{u}}-\tilde{u}m/R^2\right)_{\Sigma} .
\end{equation}
The results for the first junction condition implies that $ \tilde{R}_{\Sigma}=((r/A) \partial B/\partial t)_{\Sigma} $, and further manipulations lead us to
 $$ \tilde{u}_{\Sigma}^{-1}=(1+(r/B) \partial B/\partial r + (r/A) \partial B/\partial t)_{\Sigma} .$$
We can further differentiate $ \tilde{u}_{\Sigma} $ with respect to $\tau $ and determine $ \tilde{{\tilde{u}}}_{\Sigma} $. Using \eqref{junt}, and substituting all the previous results, and performing lengthy algebraic manipulations we finally get the following relation on $ \Sigma $:
\begin{equation}\label{jn1}
p_{\Sigma}=(qB)_{\Sigma}=(qB_{0}f)_{\Sigma}.
\end{equation}
The above relation indicates that the isotropic pressure of the interior fluid at the boundary is non-zero. The pressure only vanishes for a vanishing heat flux $q_{\Sigma}=0$ on the bounding surface $\Sigma$. This makes perfect sense as we need a non-zero heat flow at $ \Sigma $ in order to have an outgoing radiation, so that the exterior spacetime may be modelled by the Vaidya metric. This also implies that for the static case (without heat flow) no radiation can exist in the exterior, meaning that the exterior could be a Schwarzschild spacetime. In other words, at $r=r_{\Sigma}$, the isotropic pressure for the interior shear free static perfect fluid (without heat flow) vanishes, i.e., $p_{0}\vert_{\Sigma}=0$.
\\
Consequently, imposing the condition of vanishing pressure (static solution) $p_{0}\vert_{\Sigma}=0$ on the boundary surface $\Sigma$, and using equations \eqref{jn1}, \eqref{pr}, and \eqref{qa}, we obtain the following second-order differential equation for $f(t)$
\begin{equation}\label{f}
2f\ddot{f}+\dot{f}^2-2a\dot{f}=0; \;\;\;\; a=\left(\frac{A_{0}^{'}}{B_{0}}\right)_{\Sigma} ,
\end{equation}
where $ A_{0}^{'} $ represents the radial derivative $ \partial A_{0}/\partial r $. It should be noted that the constant $ a $ must be positive if the static solution generated by $A_{0}$ and $ B_{0} $ is to match with a Schwarzschild exterior.\\
The first and second integrals of the differential equation \eqref{f} are given by
\begin{eqnarray}\label{dotf}
&\frac{\dot{f}}{f}=-\frac{2a}{bf^{3/2}}+\frac{2a}{f},\\
& t-t_{s}=\frac{f}{2a}+\frac{\sqrt{f}}{ab}+\frac{1}{ab^2}\ln(1-b\sqrt{f}); \;\; b>0, \;\; b\sqrt{f}<1 ,
\end{eqnarray}
where $ b $ and $ t_{s} $ are constants of integration. It is clear that $ f $ diminishes monotonically from the value $ 1/b^2 $ at $ t=-\infty $ to $ 0 $ at $ t=t_{s} $. This also means that $ \dot{f} $ begins from $ 0 $ at $ t=-\infty $ and diminishes to $ -\infty $ at $ t=t_{s} $.
Subsequently $ \ddot{f} $ can be represented in terms of $ f $ in the following way:
\begin{equation}\label{ddotf}
\frac{\ddot{f}}{f}=-\dfrac{2a^2}{b^2 f^3}+ \dfrac{2a^2}{bf^{5/2}}.
\end{equation}
It is clear that $ \ddot{f} $ starts from $ 0 $ at $ t=-\infty $ and decreases to $ -\infty $ at $ t=t_{s} $.
The evolution of the interior is dominated by the function $ f(t) $. For the sake of physical viability it is clear that we require $\dot{f}<0 $. This would represent a contracting interior. Moreover, for the heat flow to be positive so as to ensure an outgoing radiation, $\dot{f} $ must be negative, since the condition $\dot{f}>0 $ would mean that heat is flowing inwards and the interior is expanding, resulting in a negative $ p_{\Sigma} $. This would also mean that $ dm/du >0 $, i.e., the negative energy density for the exterior spacetime is flowing out thereby causing the central mass to increase in size.
Therefore we can consider the system to represent spherical collapse from a static sphere with heat flowing outwards and leaving the sphere in the form of outgoing radiation. This is a physically reasonable system for a collapsing star with outgoing radiation, for which the arrow of time points in the direction of increasing $ t $ or $ u $.


\section{Analysis}
For the above solution we can reduce the field equations to a more compact form, from which the temporal $ (f) $ and radial dependence $(A_{0}, B_{0})$ can be observed clearly:
\begin{eqnarray}
p=\frac{p_{0}}{f^2}-\frac{4a^2}{8\pi A_{0}^2}\frac{\left(b\sqrt{f}-1\right)}{bf^{5/2}},\\
\mu=\frac{\mu_{0}}{f^2}+\dfrac{48a^2}{8\pi A_{0}^{2}b^{2}}\dfrac{\left(b\sqrt{f}-1\right)^2}{f^3},\\
q=-\dfrac{4aA_{0}^{'}}{8\pi b A_{0}^{2}B_{0}^{2}}\dfrac{\left(b\sqrt{f}-1\right)}{f^{7/2}} .
\end{eqnarray}
It is evident that for $ b\sqrt{f}=1 $ (which occurs in the limit $  t \rightarrow -\infty$), the isotropic pressure and energy density of the fluid configuration reduces to their static counterparts, i.e., $p|_{t=-\infty}=b^{4}p_{0}$, $\mu|_{t=-\infty}= b^{4}\mu_{0} $, and the heat flux vanishes in this limit ($ q|_{t=-\infty}=0 $). We find that at later times (approaching $ t_{s} $), the function $ f $ diminishes rapidly, and in this limit the matter variables behave as $ q_{a}\sim 1/f^{7/2} $, with $ \mu \sim 1/f^{3} $ and $ p\sim 1/f^{5/2} $. We can see that the heat flux is changing faster than the other variables. This might not be true in presence of shear. Naturally in the limit $ t \rightarrow t_{s} $, when $ f=0 $, all the above quantities diverge. We can also compute the total mass-energy bounded by the surface $ \Sigma $ using the equation \eqref{sigmass}:
\begin{eqnarray}
m(u)&=&\Bigg\lbrace \dfrac{2a^{2}r^{3}B_{0}^{3}}{A_{0}^{2}b^{2}}\left(-1+b\sqrt{f}\right)^{2}+\left(-r^{2}B_{0}^{'} -\frac{r^{3}B_{0}^{'2}}{2B_{0}}\right)f\Bigg\rbrace_{\Sigma}\nonumber\\
&=& \Bigg\lbrace \dfrac{2a^{2}r^{3}B_{0}^{3}}{A_{0}^{2}b^{2}}\left(-1+b\sqrt{f}\right)^{2}+ m_{0}f\Bigg\rbrace_{\Sigma},
\end{eqnarray}
where $ m_{0}\equiv \left(-r^{2}B_{0}^{'}-{r^{3}B_{0}^{'2}}/{2B_{0}}\right)\vert_{\Sigma}$, is the total energy inside $ \Sigma $ for the static system. For the sake of clarity we can also choose some appropriate radial functions as $ A_{0} $ and $ B_{0} $. We know that at $ t=-\infty $, matter inside the star is in the static configuration of a perfect fluid, which resembles the Schwarzschild interior solution \cite{ray}.\\
For the sake of simple illustration we assume the following functional forms :
\begin{eqnarray}\label{A0B0}
A_{0}=\alpha-\beta\dfrac{1-r^2}{1+r^2}, \qquad \;\;\; B_{0}=\dfrac{2\gamma}{1+r^2},
\end{eqnarray}
where $ \alpha, \beta $ and $ \gamma $ are constants. Using these functions we can further simplify the total energy trapped inside $\Sigma$ for the static system to be given by $ m_{0}=\dfrac{4\gamma r_{\Sigma}^{3}}{(1+r_{\Sigma}^{2})^{3}} $.\\
These functional forms correspond to static conformally flat spacetimes. The above result shows the radial dependence of the matter variables in an approximate manner. We cannot use them to compute the curvatures, because the Weyl curvature is simply zero in such cases, and the gravitational entropy epoch function would vanish for all times. Therefore, we need more involved and realistic functional forms to do proper calculations. We have avoided specific radial functions and have directly investigated the behaviour of the curvatures as functions of time. \\
Instead of using a timelike hypersurface $ \Sigma $ for the purpose of matching, another physically motivated option would be a smooth matching at a spacelike hypersurface $ \Xi $ marked by $t = t_{0}$ (i.e., a hypersurface defined by the equation $ f(\rho,t)=t-t_{0}=0 $), with a static spacetime for $t < t_{0}$ which becomes dynamic for $t > t_{0}$. We can write the hypersurface $ \Xi $ in terms of its intrinsic coordinates in the form
\begin{equation}
ds_{\Xi}^{2}=\chi^{2}(x,t_{0})[dx^{2}+x^{2}d\Omega^{2}].
\end{equation}
This means that until the time $ t_{0} $, the system is in its static configuration with no heat flux, and after that when the collapse begins, the interior spacetime becomes dynamic with non-zero heat flux. We can represent these two different spacetimes as
\begin{eqnarray}
ds_{dy}&=&-A^{2}(\rho,t)dt^{2}+B^{2}(\rho,t)[d\rho^{2}+\rho^{2}d\Omega^{2}], \;\;\;\; t> t_{0} \label{stdy1} \\
ds_{st}&=&-A_{0}^{2}(r)dt^{2}+B_{0}^{2}(r)[dr^{2}+r^{2}d\Omega^{2}], \;\;\;\; t\leq t_{0}, \label{stdy2}
\end{eqnarray}
where $ d\Omega^{2}= d\theta^{2}+ \sin^{2}\theta d\phi^{2}$ is the line element on a two-sphere and we will also use the conditions \eqref{int_soln}. Since the $4$-velocity is normal to the hypersurface, Darmois-Israel matching conditions require the continuity of the first and second fundamental forms at $t = t_{0}$. This gives us the following results:
$$ r=\rho=x, B_{0}(r)= \chi(x,t_{0})  .$$
For the model used by us, consisting of \eqref{stdy1} and  \eqref{stdy2}, a necessary and sufficient condition for this matching is $f(t_{0}) = 1$ for $ t \leq t_{0}$, and for the collapse to take place we need $\dot{f} < 0$, with $f(t) < 1$ for $t > t_{0}$. The condition of isotropy of pressure implies the following relation for the static fluid
\begin{equation}\label{isp}
\frac{A_{0}^{''}}{A_{0}}+\frac{B_{0}^{''}}{B_{0}}=\left(\frac{A_{0}^{'}}{A_{0}} +\frac{B_{0}^{'}}{B_{0}}\right)\left(\frac{1}{r}+2\frac{B_{0}^{'}}{B_{0}}\right).
\end{equation}
We can choose the radial functions $A_{0}(r)$ and $B_{0}(r)$ in such a manner that they satisfy the constraint \eqref{isp}. Since $t_{0}$ can provide us the initial data for $t > t_{0}$, this matching will help us to verify whether the ratios $C_{abcd}C^{abcd}(R_{ab}R^{ab})^{-1}$ and $C_{abcd}C^{abcd}(R_{abcd}R^{abcd})^{-1}$ are well behaved functions of $r$, or not. We have already chosen such functions in \eqref{A0B0}, which satisfies \eqref{isp}. As discussed earlier, these functions correspond to conformally flat Schwarzschild interior solution \cite{ray}, and therefore we cannot evaluate the GE function using them. Finding a correct form of these functions and performing the necessary calculations is a task in itself, which we are not considering here. Some realistic functional forms for $ A_{0}$ and $ B_{0} $ can be found in \cite{gold1,gold2}. \\
To begin our analysis of gravitational entropy we need to find the curvature scalars for the interior spacetime consisting of a collapsing sphere with radial heat flow. We find that the Ricci square is given by the following expression
\begin{equation}\label{risq}
R_{ab}R^{ab}=\dfrac{64\pi Y}{f^4},
\end{equation}
where the term $ Y $ is given by the expression
\begin{equation}\label{y}
Y\equiv \left(\mu_{0}+\dfrac{3{\dot{f}}^2}{8\pi A_{0}^2}\right)^2+3\left(p_{0}^2-\dfrac{4p_{0}a\dot{f}}{8\pi A_{0}^2} \right)+\left(\dfrac{\dot{f}}{4\pi A_{0}^2}\right)^2 \left[ 3a^2-2\left(\frac{A_{0}^{'}}{B_{0}}\right)^2    \right].
\end{equation}
\\
The above expression of $ Y $ can be further simplified by substituting $ \dot{f} $ in terms of $ f $ by using \eqref{dotf}, which yields
\begin{eqnarray}\label{Ybf}
Y & =\left(\mu_{0}^{2}+3p_{0}^2\right)+\frac{9a^4}{4\pi^{2}A_{0}^{4}}\left(1-\frac{1}{b\sqrt{f}}\right)^{4}-\dfrac{3p_{0}a^{2}}{\pi A_{0}^{2}}\left(1-\frac{1}{b\sqrt{f}}\right) \nonumber\\
& + 4a^{2}\left[ \dfrac{3\mu_{0}}{4\pi A_{0}^{2}}+\dfrac{\bigg\lbrace 3a^{2}-2\left(\frac{A_{0}^{'}}{B_{0}}\right)^{2}\bigg\rbrace}{16\pi^{2} A_{0}^{4}} \right]\left(1-\frac{1}{b\sqrt{f}}\right)^{2} .
\end{eqnarray}
Therefore for all values of $ r $ such that $3a^2 \geq 2\left({A_{0}^{'}}/{B_{0}}\right)^2  $ (this includes the vicinity of $ \Sigma $ and the centre of the sphere at $ r=0 $ where $A_{0}^{'}$ vanishes), the value of $ Y $ will increase with time. If we now investigate the dependence of $ Y $ on time, we see that at $ t=-\infty $, i.e., when $ b\sqrt{f}=1 $, the value of $ Y $ becomes $ Y|_{ t=-\infty }=(\mu_{0}^{2}+3p_{0}^{2}) $. As time passes (i.e., $ b\sqrt{f}<1 $), the value of $ Y $ increases such that $ Y> (\mu_{0}^{2}+3p_{0}^{2})$ at all times. Finally at $ t=t_{s} $, when $ b\sqrt{f}=0 $, the value of $ Y $ diverges to $ +\infty $. Here $ \dot{f} $ must be negative for a physically viable collapse as discussed in the previous section. This makes $R_{ab}R^{ab} $ an increasing function, starting from $R_{ab}R^{ab}\vert_{t=-\infty}=64\pi b^{8}(\mu_{0}^{2}+3p_{0}^{2})$  and diverging to $ +\infty  $ at $ t=t_{s} $.
We will also need the field equations of the static solution (which are given below), and will obtain the energy density and pressure as functions of $A_{0}$ and $B_{0}$ in order to express the necessary terms in a convenient way. We therefore have\\
\begin{equation}\label{mu0}
8\pi \mu_{0}=-\dfrac{1}{B_{0}^2}\left[2\frac{B_{0}^{''}}{B_{0}}-\left(\frac{B_{0}^{'}}{B_{0}}\right)^2+\frac{4}{r}\frac{B_{0}^{'}}{B_{0}}\right]=\dfrac{3}{\gamma^2},
\end{equation}
and
\begin{eqnarray}\label{p0}
8\pi p_{0} &=& \dfrac{1}{B_{0}^2}\left[ \left(\frac{B_{0}^{'}}{B_{0}}\right)^2+ \frac{2}{r}\frac{B_{0}^{'}}{B_{0}}+2\frac{A_{0}^{'}}{A_{0}}\frac{B_{0}^{'}}{B_{0}}+\frac{2}{r}\frac{A_{0}^{'}}{A_{0}}\right] \nonumber\\
&=& \frac{1}{\gamma^2}\left[-1+\dfrac{2\beta (1-r^{2})}{\alpha(1+r^2)-\beta(1-r^2)}\right].
\end{eqnarray}
\\
The junction condition for the static solution, i.e., $ p_{0}\vert_{\Sigma}=0 $, leads us to the relation
\begin{equation}
\frac{\alpha}{\beta}=3\dfrac{1-r_{\Sigma}^2}{1+r_{\Sigma}^2}.
\end{equation}
We now assume $ \beta=1/2 $, and we also find that $ a={r_{\Sigma}}/{\gamma (1+r_{\Sigma}^{2})} $. Let us now write down the field equations in terms of \eqref{A0B0}, so that we can clearly observe the radial dependence:
\begin{eqnarray}
p=\frac{p_{0}}{f^2}-\frac{1}{8\pi}
\Bigg\lbrace \dfrac{16r_{\Sigma}^{2}(1+r^{2})^{2} }{\gamma^{2}\lbrace 3(1-r_{\Sigma}^{2})(1+r^{2})-(1+r_{\Sigma}^{2})(1-r^{2})\rbrace^{2}}  \Bigg\rbrace
\frac{\left(b\sqrt{f}-1\right)}{bf^{5/2}},\\
\mu=\frac{\mu_{0}}{f^2}+\dfrac{3}{8\pi}
\Bigg\lbrace \dfrac{16r_{\Sigma}^{2}(1+r^{2})^{2} }{\gamma^{2}\lbrace 3(1-r_{\Sigma}^{2})(1+r^{2})-(1+r_{\Sigma}^{2})(1-r^{2})\rbrace^{2}}  \Bigg\rbrace
\dfrac{\left(b\sqrt{f}-1\right)^2}{f^3},\\
q=-\dfrac{1}{8\pi}
\Bigg\lbrace \dfrac{8rr_{\Sigma}(1+r^{2})^{2}(1+r_{\Sigma}^{2}) }{\gamma^{3}\lbrace 3(1-r_{\Sigma}^{2})(1+r^{2})-(1+r_{\Sigma}^{2})(1-r^{2})\rbrace^{2}}  \Bigg\rbrace
\dfrac{\left(b\sqrt{f}-1\right)}{f^{7/2}}.
\end{eqnarray}
Here $ \mu_{0} \, \textrm{and} \, p_{0} $ must be positive for them to be physically viable. Similarly, $ m(u) $ can be computed as
\begin{equation}
m(u)= \Bigg\lbrace \dfrac{16 \gamma r_{\Sigma}^{5}}{(1+r_{\Sigma}^{2})^{3}(1-r_{\Sigma}^{2})^{2}} \Bigg\rbrace
\dfrac{\left(-1+b\sqrt{f}\right)^{2}}{b^{2}}+\dfrac{4\gamma r_{\Sigma}^{3}}{(1+r_{\Sigma}^{2})^{3}} .
\end{equation}
Subsequently, the following expression of the Weyl curvature scalar can be deduced from the metric \eqref{int}:
\begin{eqnarray}\label{weyl1}
&& C_{abcd}C^{abcd} \nonumber \\
  &=& \dfrac{4}{3}\dfrac{1}{A_{0}^2 B_{0}^8 f^4 r^2} [A_{0}^{''}B_{0}^2 r -B_{0}^{''}A_{0}B_{0}r-2A_{0}^{'}B_{0}B_{0}^{'}r+2A_{0}{B_{0}^{'}}^2 r- \nonumber \\ 
&  &A_{0}^{'}B_{0}^2+ A_{0} B_{0} B_{0}^{'}]^2   \nonumber \\
& = &  \dfrac{4}{3B_{0}^4 f^{4}r^2}\bigg[\frac{A_{0}^{''}}{A_{0}}r-\frac{B_{0}^{''}}{B_{0}}r-2\frac{A_{0}^{'}}{A_{0}}\frac{B_{0}^{'}}{B_{0}}r + 2\bigg(\frac{B_{0}^{'}}{B_{0}}\bigg)^{2}r-\frac{A_{0}^{'}}{A_{0}}+\frac{B_{0}^{'}}{B_{0}}\bigg]^{2} .
\end{eqnarray}
Now, using the relation \eqref{isp} we obtain the following after simplification:
\begin{equation}\label{ispr}
\frac{A_{0}^{''}}{A_{0}}r-\frac{B_{0}^{''}}{B_{0}}r-2\frac{A_{0}^{'}}{A_{0}}\frac{B_{0}^{'}}{B_{0}}r + 2\bigg(\frac{B_{0}^{'}}{B_{0}}\bigg)^{2}r-\frac{A_{0}^{'}}{A_{0}}+\frac{B_{0}^{'}}{B_{0}} =2\bigg[\frac{B_{0}^{'}}{B_{0}}+2\bigg(\frac{B_{0}^{'}}{B_{0}}\bigg)^{2}r-\frac{B_{0}^{''}}{B_{0}}r\bigg] .
\end{equation}
Using \eqref{ispr} we then simplify the Weyl square in \eqref{weyl1} into the compact form
\begin{equation}
C_{abcd}C^{abcd}=\dfrac{16}{3B_{0}^{4}f^{4}r^{2}}\bigg[\frac{B_{0}^{'}}{B_{0}} +2\bigg(\frac{B_{0}^{'}}{B_{0}}\bigg)^{2}r-\frac{B_{0}^{''}}{B_{0}}r\bigg]^{2}=\dfrac{16}{3}(B_{0}f)^{-4}X^2,
\end{equation}
where $ X $ is given by the expression
\begin{equation} \label{X}
X\equiv \left( \frac{B_{0}^{''}}{B_{0}}-2\frac{{B_{0}^{'}}^{2}}{B_{0}^{2}} -\frac{B_{0}^{'}}{rB_{0}}  \right) .
\end{equation}
Thus the spacetime becomes conformally flat for all times if $ X=0 $. In such a scenario neither $ P $ nor $ P_{1} $ can provide us any indication of the arrow of time.
If we assume specific functional forms of $ A_{0} $ and $ B_{0} $ by hand, as is given in \eqref{A0B0}, we will find that the quantity $ X $ vanishes. Therefore these choices are not suitable for curvature calculations. We will need more involved and realistic functional forms to do curvature calculations. If we can choose some other functional form of $ A_{0} $ and $ B_{0} $ which yields a non-zero $ X $, such that the Weyl scalar will always be positive, then we can always calculate the gravitational epoch function $ P $ defined in \eqref{P} from the relation
\begin{equation}
P=\dfrac{X^2}{12 \pi Y B_{0}^{4}}.
\end{equation}
Here as time passes, the value of $ Y $ will increase, and $ P $ will decrease with time. To begin with, for $ t=-\infty $ we will have
\begin{equation}
P|_{t=-\infty}=\dfrac{X^{2}}{12 \pi B_{0}^{4}(\mu_{0}^{2}+3p_{0}^{2})},
\end{equation}
making $ P $ a positive function in this limit, and at $ t=t_{s} $ the value of $ P $ will vanish. Therefore the gravitational entropy would decrease with time, which is what Bonnor obtained in \cite{Bonnor3}. We will now proceed to calculate the gravitational epoch function $ P_{1} $.
\\
We can write the expression for the Kretschmann scalar in terms of other curvature scalar quantities to obtain
\begin{align}\label{kr1}
R_{abcd}R^{abcd} &= C_{abcd}C^{abcd}+2R_{ab}R^{ab}-\dfrac{1}{3}R^2 \nonumber\\
&=\dfrac{16}{3}\dfrac{X^2}{(B_{0}f)^4}+\dfrac{128\pi Y}{f^4}-\frac{1}{3}R^2.
\end{align}
Next, we determine the Ricci scalar $ R $ in order to evaluate the full expression for the Kretschmann scalar $ R_{abcd}R^{abcd} $. From the interior metric we can determine the following expression for the Ricci scalar as
\begin{eqnarray}
  R &=& \dfrac{1}{B_{0}^4 f^2 r A_{0}^2}[-2A_{0}^{''} A_{0}B_{0}^2 r - 4B_{0}^{''}A_{0}^2 B_{0}r + 6 \ddot{f}B_{0}^4 fr + 2 {B_{0}^{'}}^2 A_{0}^2 r ] \nonumber \\
  & & + \dfrac{1}{B_{0}^4 f^2 r A_{0}^2}[ -2 A_{0} B_{0}(r A_{0}^{'}+4A_{0})B_{0}^{'}+6\dot{f}^2 B_{0}^4 r- 4 A_{0}^{'}A_{0}B_{0}^2]. \label{ricci1}
\end{eqnarray}
\\
Using the three relations \eqref{mu0}, \eqref{p0}, and \eqref{isp}, along with \eqref{ricci1}, we get the final simplified expression of the Ricci scalar given by the following

\begin{eqnarray}\label{ricci2}
R &=& \dfrac{6}{f^2 A_{0}^2}(f\ddot{f}+\dot{f}^2)+\dfrac{8\pi}{f^2}(\mu_{0}-3p_{0})\\
  &= & \dfrac{6}{f^2 A_{0}^2}\left[2a^{2}\left(\frac{1}{b\sqrt{f}}-1\right)\left(\frac{1}{b\sqrt{f}}-2\right)\right]+\frac{8 \pi}{f^{2}}(\mu_{0}-3p_{0}) .
\end{eqnarray}
From the above expression of the Ricci scalar it is clear that $ R|_{t=-\infty}= 8\pi b^{4}(\mu_{0}-3 p_{0}) $, and at $ t=t_{s} $ the scalar curvature diverges to $ +\infty $. Therefore the Ricci scalar is increasing with time, which is also evident from the functional form given above.
Finally we obtain the expression \eqref{kr2} for the Kretschmann scalar by putting \eqref{ricci2} in \eqref{kr1}
\begin{equation}\label{kr2}
R_{abcd}R^{abcd}=\frac{1}{f^4}\left[\dfrac{16}{3}\dfrac{X^2}{(B_{0})^4}+128\pi Y-\frac{1}{3}\left(\dfrac{6}{ A_{0}^2}(f\ddot{f}+\dot{f}^2)+8\pi(\mu_{0}-3p_{0})\right)^2\right].
\end{equation}
We can now use the expression of $ Y $ from \eqref{Ybf} and substitute $ \dot{f}, \ddot{f} $ using \eqref{dotf}, \eqref{ddotf} to get the Kretschmann scalar in terms of $ f $.
Now taking the limit $ t= -\infty $ and also using \eqref{f}, the Kretschmann scalar in this limit becomes
\begin{equation}
R_{abcd}R^{abcd}\vert_{t=-\infty}=\frac{1}{f^4}\left[\dfrac{16}{3}\dfrac{X^2}{B_{0}^4}+128\pi (\mu_{0}^2 + 3 p_{0}^2)-\frac{1}{3}64\pi^2(\mu_{0}-3p_{0})^2\right].
\end{equation}
\\
Finally we can evaluate the gravitational entropy epoch function $ P_{1} $ at the limit $ t=-\infty $ as
\begin{eqnarray}
P_{1}\vert_{t=-\infty} &=& (C_{abcd}C^{abcd})(R_{abcd}R^{abcd})^{-1}\vert_{t=-\infty}\nonumber \\
&=& \dfrac{16X^2}{16 X^{2}+384\pi B_{0}^{4}(\mu_{0}^{2}+3p_{0}^{2})-64 \pi^{2} B_{0}^{4}(\mu_{0}-3p_{0})^{2}} .
\end{eqnarray}

Therefore the quantity $ P_1$ is positive in this limit, to begin with. Now let us consider the other limit when $ t\rightarrow t_{s}$ \cite{Bonnor3}. In this limit, the Kretschmann scalar goes to infinity. As a result, in this limit we find that $ P_1 \rightarrow 0 $. Here $ t_{s} $ is the constant of integration which comes from the second integral of the equation \eqref{f}. Without any hesitation we can conclude that the gravitational entropy given by the Rudjord proposal decreases with time in this scenario. However we know that the radiation (thermodynamic) arrow should point in the direction of increasing $ t $. So, for physical viability, the gravitational epoch function $ P_{1} $ should increase with time, contrary to what we obtained. Therefore the gravitational arrow of time defined by the gravitational epoch function $ P_{1} $ gives us a wrong sense of time, a result similar to what Bonnor obtained \cite{Bonnor3} with the epoch function $ P $.

\section{Conclusions}

In this paper, we have examined the status of the gravitational arrow of time during the spherical collapse of radiating fluid which conducts heat. Specifically, we checked whether the results obtained by W. B. Bonnor \cite{Bonnor1} where he found the gravitational arrow of time to be opposite to the thermodynamic arrow of time, could be confronted with a different choice of the gravitational epoch function. The measure of gravitational epoch function $P$ used by Bonnor was given by the ratio of the Weyl square to the Ricci square. Here we assumed the gravitational epoch function (or GE) $P_{1}$ to be given by the ratio of the Weyl scalar to the Kretschmann scalar. Our analysis validates Bonnor's findings, i.e., the gravitational arrow of time and the thermodynamic arrow of time point in opposite directions, even if we choose the function $P_1$.\\
It is to be noted that the epoch function $P$ vanishes for conformally flat spacetimes, and is not defined in vacuum or in presence of null fluid alone. It may provide a satisfactory orientation for the gravitational arrow of time in the case of inhomogeneous cosmological models. However, the epoch function $P_1$ has a slight edge over $P$, as it may be defined for null fluids. From that point of view, the result obtained in this paper is more generalized than that obtained by Bonnor. \\
From the above analysis it appears that the Weyl proposal of gravitational entropy is applicable only to the universe as a whole (provided that we exclude the white holes), at least for a local gravitational collapse both the proposals of GE gives a wrong sense of time, questioning the validity of a local application of the proposal of GE in the context of the arrow of time. This might be due to the shear-free model considered in both the studies. Studies on more realistic models with shear can shed further light on this issue.

\section*{Acknowledgements}
SC is thankful to CSIR, Government of India for research grant. SDM acknowledges the support provided by the South African Research Chair Initiative of the Department of Science and Technology and the National Research Foundation (NRF), South Africa. SG acknowledges IUCAA, India for an associateship and CSIR, Government of India for supporting a major research project. RG thanks NRF, South Africa, for support. The authors thank the anonymous reviewers for providing concrete suggestions towards improving the manuscript.

\section*{Data availability statement}
No new data were created or analysed in this study.

\end{document}